# Full Two-Port S-Parameters at mK Temperatures: a Calibration Strategy and Uncertainty Budget

Luca Oberto, Ehsan Shokrolahzade, Emanuele Enrico, Luca Fasolo, Andrea Celotto, Bernardo Galvano, Alessandro Alocco, Paolo Terzi, Faisal A. Mubarak, *Senior Member, IEEE* and Marco Spirito, *Member, IEEE*

*Abstract*—**This paper describes the developed setup and characterization approach for full two-port calibrated S-parameter measurements at cryogenic temperatures, together with a complete uncertainty budget. The system developed at the Istituto Nazionale di Ricerca Metrologica (INRiM, Italy), exploits the Short-Open-Load-Reciprocal technique to realize error-corrected cryogenic measurements within a single cooling cycle. The system operates down to mK temperatures over the 4-12 GHz band in coaxial lines. Calibration standards are referenced to SI-traceable room-temperature measurements, while a numerical approach is used to evaluate the response shift of the artifacts upon cooling and to derive an additional cryogenic uncertainty contribution for the measurement uncertainty budget. Moreover, relevant measurement uncertainty contributions are evaluated according to internationally agreed procedures, and a comprehensive uncertainty budget is presented. Test measurements on a 20 dB attenuator are shown as an example. An attenuation of 20.70 ± 0.08 dB (95% confidence interval) was obtained at 6 GHz. Full SI-traceable verification at mK temperatures remains an open challenge; however, an initial calibration verification is also presented.**

Manuscript received xxxxxxxx yy, zzzz; revised xxxxxxxx yy, zzzz; accepted xxxxxxxx yy, zzzz. Date of publication xxxxxxxx yy, zzzz; date of current version xxxxxxxx yy, zzzz. The associate editor coordinating the review process was Dr. XXX YYY.

This work was supported by the European project SuperQuant. This project 20FUN07 SuperQuant has received funding from the EMPIR programme co-financed by the Participating States and from the European Union's Horizon 2020 research and innovation programme. This work was also supported by the Italian project PRIN 2022 project CalQuStates. CalQuStates received funding by the European Union – Next Generation EU, Mission 4, Component 1, CUP E53D23002210006 *(Corresponding author: Luca Oberto).*

L. Oberto, E. Enrico, L. Fasolo and P. Terzi are with Istituto Nazionale di Ricerca Metrologica, Torino, 10135 Italy (e-mail: l.oberto@inrim.it, e.enrico@inrim.it, l.fasolo@inrim.it, p.terzi@inrim.it).

E. Shokrolahzade and M. Spirito are with Delft University of Technology, Delft, 2628 CD The Netherlands (e-mail: e.shokrolahzade@tudelft.nl, m.spirito@tudelft.nl).

A. Celotto and A. Alocco are with Istituto Nazionale di Ricerca Metrologica, Torino, 10135 Italy, and with Politecnico di Torino, Torino, 10129 Italy (e-mail: andrea.celotto@polito.it, alessandro.alocco@polito.it).

B. Galvano is with Istituto Nazionale di Ricerca Metrologica, Torino, 10135 Italy, with University of Palermo, Palermo, 90133 Italy, and with Consorzio Nazionale Interuniversitario per le Telecomunicazioni, Parma, 43124 Italy (e-mail: bernardo.galvano@you.unipa.it).

F. A. Mubarak is with Delft University of Technology, Delft, 2628 CD The Netherlands and with National Metrology Institute of the Netherlands – VSL, Delft, 2629 JA The Netherlands (e-mail: FMubarak@vsl.nl).

## I. INTRODUCTION

Driven by global quantum technology initiatives, companies are developing custom components for cryogenic temperatures to operate in various applications, many of them working at Microwave (MW) frequencies. However, existing MW calibration capabilities and traceability paths are based on room temperature (RT), lacking accuracy benchmarks for cryogenic applications. The absence of primary standards and traceability at cryogenic temperatures poses a challenge. The EMPIR SuperQuant project [1] has been working to address this gap by establishing innovative metrological tools for MW measurements in cryogenic environments down to mK temperatures. A collaborative effort between Istituto Nazionale di Ricerca Metrologica (INRiM) and Delft University of Technology targets the design and implementation of a cryogenic scattering (S-) parameters measurement system.

The precise measurement of S-parameters at mK temperatures is a fundamental requirement for the development and scaling of superconducting quantum information systems [2]. Since microwave components are often designed for RT use, their behavior can change significantly upon cooling. Thorough characterization allows for the de-embedding of these intervening components, moving the measurement reference plane directly to the Device Under Test (DUT) allowing its accurate calibration.

While previous work at National Institute for Standards and Technology (NIST, USA) pioneered the evaluation of full two-port S-parameters at MW frequencies in such extreme temperature range through an adapted thru-reflect-line (TRL) algorithm [2], the field has since expanded to include a variety of robust *in-situ* methodologies. Later studies demonstrated the effectiveness of databased one-port short-open-load (SOL) calibration using commercial standards to characterize qubit drive line components and superconducting resonators [3] – [4]. TRL schemes based on custom coplanar waveguide (CPW) or grouped CPW standards, enabled the de-embedding of non-connectorized quantum integrated circuits [5]. The short-open-load-reciprocal (SOLR, also known as Unknown Thru) technique [6], has emerged as a convenient method for full two-port and even multiport characterization, as it avoids the need for a precisely defined Thru standard [7] – [9]. Unlike [8], the present work does not investigate a wide temperature range, but is specifically focused on the mK regime, which is particularly relevant for the operation and characterization of superconducting quantum devices and microwave circuits [3]. Moreover, unlike [9], the present implementation is limited to two-port measurements.

Contrary to earlier assumptions that calibration standards remain stable across extreme temperature gradients from RT to mK, recent literature has begun to quantify the temperature-dependent behavior of impedance standards [8]. Studies have shown that, while zero-offset shorts are relatively stable, the microwave response of load standards can deviate significantly due to changes in resistivity [3], [8]. Despite these findings, an approach to calibration standards definition that could preserve SI traceability remains absent.

Moreover, initial estimates of cryogenic S-parameter uncertainty identified variability and asymmetry in the switch devices employed to toggle between calibration standards and DUTs, and in jumper cables used to connect calibration standards to the switch ports as dominant error sources [2]. Recent studies evidenced that the contribution of calibration standards is also of major importance [4], [9]. However, a comprehensive uncertainty budget remains a critical gap in the literature. Explicit traceability-chain and uncertainty-propagation frameworks for S-parameter measurements have also been discussed in the literature [10], although not for the present cryogenic mK case.

To address these limitations, this work leverages databased methodologies, wherein suitable commercial standards are employed [3]. After a SI traceable pre-characterization at RT, the change of their characteristics at mK temperatures is evaluated by means of SI traceable DC measurements and mechanical specifications, and electro-thermal simulations. When the data employed in the simulation environment are obtained from measurements or other sources maintaining SI traceability, the resulting model response is a best effort to preserve that traceability. Moreover, the shift in response from RT to mK can be employed as an uncertainty expansion that preserves a link to traceable data.

By implementing the SOLR technique, we maintain broadband behavior while minimizing the physical volume of calibration components. Furthermore, we exploit the direct access to Vector Network Analyzer (VNA) receivers, thereby improving dynamic range. To the best of the authors' knowledge, we also present the first comprehensive uncertainty budget at mK temperatures, including a thorough evaluation of all relevant contributions. Moreover, although full verification of S-parameter measurements in coaxial lines at cryogenic temperatures remains an open challenge, we perform an initial SI-traceable verification of one-port measurements against known standards. In addition, the response symmetry of a reciprocal DUT is used as a qualitative plausibility test for transmission measurements.

Following this approach, we consider the resulting measurements to retain at least partial traceability to the SI.

As an example, a 20 dB attenuator is used as a DUT. Finally, a comparison is made between RT and cryogenic measurements, as well as between cryogenic SOLR calibration and complex Thru normalization, demonstrating the advantage of the proposed approach. This extends the work previously presented at CPEM 2024 [11].

In the following section, the cryogenic measurement set-up is described in section II. The choice and characterization of the calibration standards are presented in section III, while section IV highlights the procedure for VNA calibration and the evaluation of the measurement uncertainty. Measurement results are presented in section V. Moreover, calibration verification is discussed in section VI. Finally, conclusions are drawn in section VII.

## II. Cryogenic S-parameters Measurement Set-up

The design concept of the measurement system aimed to adapt conventional S-parameter methods for operation within a dilution refrigerator. A simplified schematic of the realized measurement setup [11] is presented in Fig. 1 (showing only key elements for space consideration). It utilizes two heavily attenuated coaxial input lines (circa 55 dB in total at RT, including three 10 dB attenuators – only one of which is visible in the picture –, cables and -10 dB directional couplers) and two superconducting output lines (blue lines in Fig. 1).

The high attenuation of the input lines allows reduction of the thermal noise at the input port of the DUT below the single photon level, without saturating the cooling power of the refrigerator [12]. This is particularly useful when characterizing quantum devices such as those needed for quantum computing or other quantum technologies. With the configuration depicted in Fig. 1, showing a total attenuation of the input lines of 55 dB (not all attenuators are shown in the picture for the sake of space), the noise photon occupation number, a dimensionless quantity that can be thought of as photon flux spectral density (i.e., as number of photons per Hz frequency interval, per second) at the cryostat mixing chamber (MXC), can be written in explicit form as a sum of the contributions of all attenuation stages [12]. In our refrigerator, attenuation stages are located at RT, 50 K, and 4 K plates, Still, Cold Plate, and MXC. Denoting by $A_m$ the linear attenuation factor of the attenuator anchored at stage $m$, and by $T_m$ its temperature, the total attenuation from RT to the MXC is $A_{tot} = \prod_{m=1}^{N} A_m$, $N$ being the number of attenuation stages. The noise photon occupation number at the MXC can then be expressed as:

$$n_{MXC}(f) = \frac{n_{\mathrm{in}}(f)}{A_{tot}} + \sum_{j=1}^{N} n_{add,j}(f) \left[\prod_{m=j+1}^{N} A_m\right]^{-1}, \quad (1)$$

in which $n_{\mathrm{in}}(f)$ is the noise photon occupation number at the input of the line (typically at RT), and $n_{\mathrm{add,j}}(f)$ is the noise contribution added by the attenuator at stage $j$. For an attenuator in thermal equilibrium at temperature $T_j$, the added noise is given by:

$$n_{add,j}(f) = \left(1 - \frac{1}{A_j}\right) \frac{1}{exp\left(\frac{hf}{k_B T_j}\right) - 1}, \quad (2)$$

where $h$ is Planck's constant, and $k_B$ is Boltzmann's constant. At $f$ = 10 GHz and with such attenuation, we estimate $n_{MXC}$ to be lower than $2 \cdot 10^{-2}$. The corresponding input noise spectral density at the DUTs $S(f) = \left(n_{MXC} + \frac{1}{2}\right) hf$ can be assumed to be dominated by vacuum fluctuations, given $n_{MXC} \ll 1$.

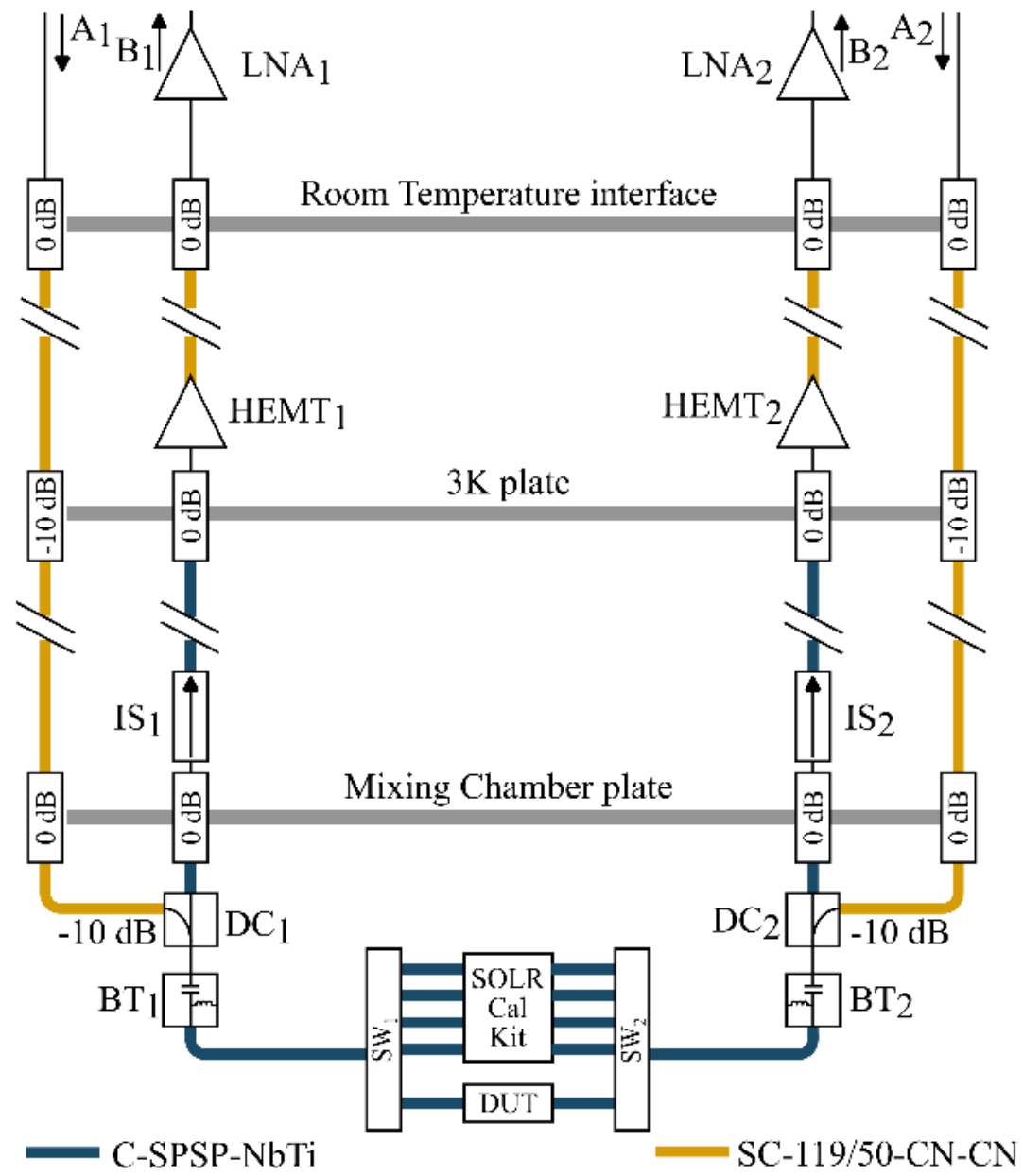

**Fig. 1.** Schematic of the INRIM cryogenic S-parameters measurement setup in which 0 dB blocks are simple SMA adaptors, ISs are 60 dB isolators, DCs are directional couplers, BT are bias-tees, SWs are electromechanical SP6T switches.

A 36 dB HEMT amplifier on the 3 K stage and a 26 dB Low Noise Amplifier (LNA) at RT are added to each output line. The gain of the amplifiers together with the attenuation of the output lines (about 8.5 dB including isolators and directional coupler insertion loss) approximately counterbalance the input line attenuation. Furthermore, two cryogenic coaxial SP6T electromechanical switches are employed to toggle between calibration standards and DUTs. Microwave signals are generated and acquired by means of a VNA. The system was mounted in a dilution refrigerator CF-CS110-500 from Leiden Cryogenics, located in the Quantum Circuit for Metrology Laboratory at INRiM. The signal separation made inside the measurement chamber by means of the directional couplers $DC_1$ and $DC_2$ allows to properly isolate input and output lines. The INRiM implementation exploits the direct receiver access of the VNA, unlike the approach in [2], which eliminates the need for RT switches that would otherwise be required to properly convey the A and B signals to the two VNA ports, further enhancing measurement repeatability and the system's dynamic range. $BT_1$ and $BT_2$ are bias-tees required to supply the bias currents, a control parameter for many quantum microwave devices whose characteristics can be tuned by means of a DC polarization, such as Josephson Traveling Wave Parametric Amplifiers (JTWPA) [13] – [15], and whose characterization relies on S-parameters measurements. The DUT is connected to the switch ports by means of CuNi/CuNi coaxial cable about 12 cm long, 1.19 mm outer diameter, providing 1.5 dB of nominal attenuation at 10 GHz.

The measurement system operates between 4-12 GHz (set by the isolators response), while the operating temperature is as low as 45 mK.

## III. Calibration Standards

The SMA artifacts to realize the Short, Open and Load calibration devices are provided by XMA Corporation-Omni Spectra [16]. As already pointed out in [3], [5], [8], dimensional metrology of the standards, and modeling of the thermal contractions are needed for high accuracy measurements. In our approach, an accurate 3D model representation of the devices is created, employing manufacturing data, in a full wave electromagnetic simulation environment, i.e., CST Studio Suite from Dassault Systemes [17]. The electrical parameters of the material used in the simulator are mapped versus temperature based on their nominal response. Given that the load resistance can exhibit a specific residual-resistivity ratio between ambient and cryogenic temperature, a direct (DC) measurement was carried out to characterize its response (see Section III.A).

The reciprocal (i.e., Thru) device is realized by a section of CuNi/CuNi coaxial cable of the same type of that used to connect the DUT to the switch ports.

Short, Open and Load standards underwent SI traceable S-parameters calibration at room temperature (RT).

### A. *Load resistance cryogenic measurement*

The resistance of the Load component was measured from RT to mK temperatures by means of a calibrated AC resistance bridge (excitation frequency 13.7 Hz), and a 4-wires technique for different dissipated powers in the range from -93 to -53 dBm. It was beforehand repeatedly immersed in liquid Nitrogen to stabilize its resistance with respect to extreme thermal cycles. Fig. 2 shows the measurements in the lowest temperature range (45 mK to 1.6 K). As it can be seen, increasing power results in a change in the resistance value, due to self-heating. From this, the thermal conductance of the component can be roughly estimated as in the order of ~pW/K.

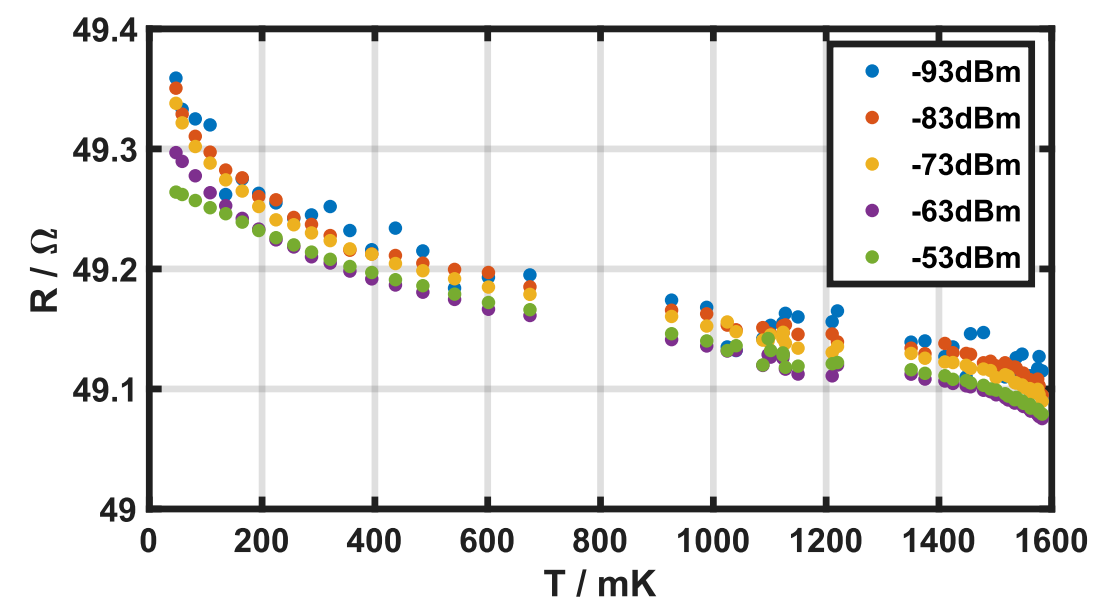

**Fig. 2.** Resistance versus temperature of the load standard for different power levels

### *B. Calibration standards behavior at RT*

The EM response of a 3D model of the standard artifacts is then benchmarked at RT by direct comparison with traceable measurement performed following the approach presented in [18]. This comparison highlighted an accurate model-to-hardware correlation as shown in Fig. 3. In the picture, the solid lines and the corresponding colored uncertainty areas represent the measurements at RT, while the dashed lines represent the outcome of the 3D EM model. As can be seen, the model agrees very well with the measurements.

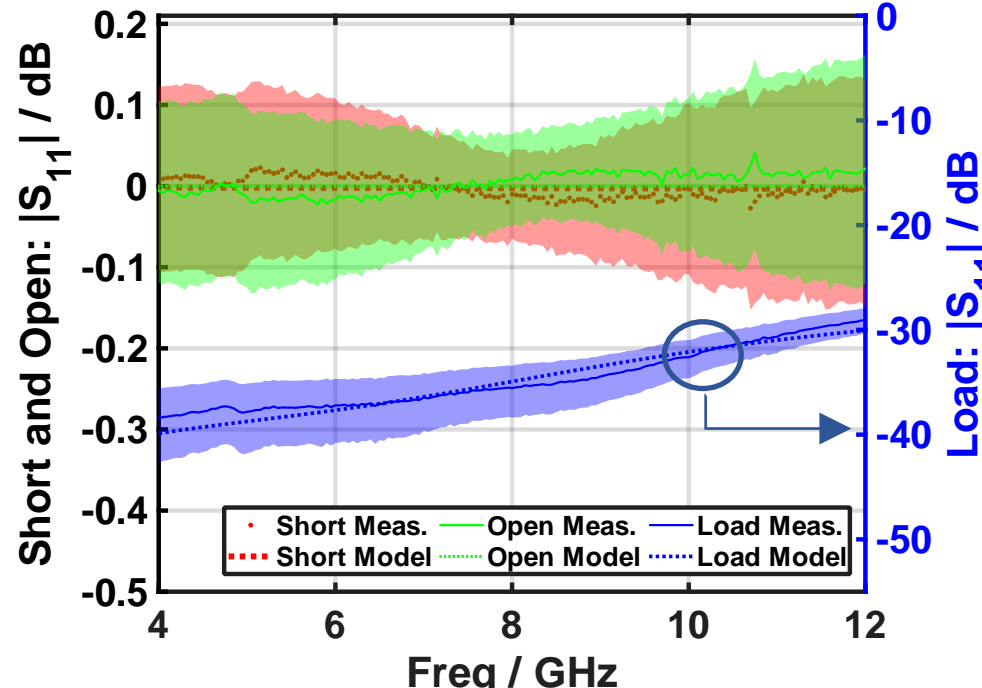

**Fig. 3.** Behavior of the 3D EM model of the Short, Open and Load standards with respect to actual RT measurements.

### *C. Evaluation of response shift at mK*

To evaluate the shift in response of the artifacts used in the calibration process when cooled at cryogenic temperatures, CST MPhysic studio is employed, to perform thermal and mechanical simulations. The various components are then simulated for their mechanical contraction when exposed to a steady state temperature of 45 mK (see Fig. 4). This new mechanical structure is then simulated for its EM response to evaluate the reflection coefficient shift with respect to RT measurements. Simulations are carried out according to the method developed in [19].

From these simulations, an uncertainty contribution accounting for the change in the behavior of the calibration standards at cryogenic temperature is extracted by calculating the complex difference between RT and mK models. It is considered to have a rectangular distribution and is added to the uncertainty budget of the calibration standards as an extra frequency-dependent uncertainty component. This uncertainty contribution is very small for Open and Short standards as shown in Table I in which, at the example frequency of 6 GHz, the overall expanded uncertainty at 45 mK ($U_{cryo}$) and at RT ($U_{RT}$) are listed in linear units for comparison, with a confidence interval of 95%.

For the Load standards, the uncertainty due to the temperature change is the dominant term, mainly due to the change in the DC resistance. This becomes clear also by looking at Fig. 5: the calibration standard providing the largest variation to the temperature shift is the Load, confirming the findings in [3], [8]. In the picture, the modulus difference is expressed in dB for better visibility.

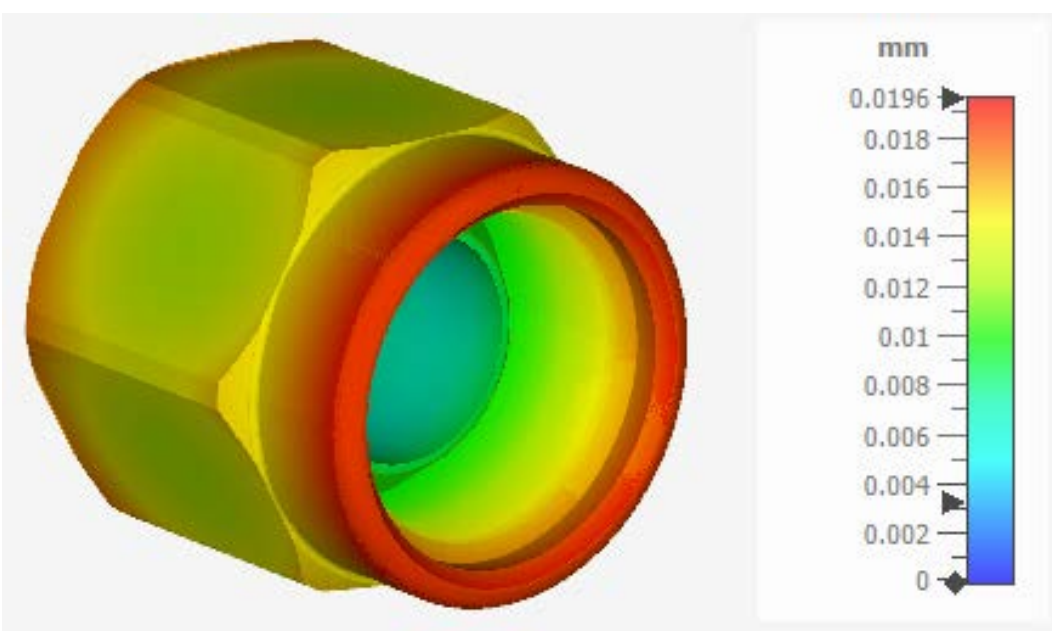

**Fig. 4.** Absolute geometrical contraction of the open standard device due to thermal shift from 293.15 K (RT) to 45 mK relative to device center.

TABLE I
COMPARISON OF THE CALIBRATION STANDARDS $|S_{11}|$ OVERALL UNCERTAINTY AT 6 GHZ AT 45 MK AND AT RT

| Standards | $U_{cryo}$ | $U_{RT}$ |
|---|---|---|
| Short | 0.003289 | 0.003287 |
| Open | 0.003975 | 0.003974 |
| Load | 0.012566 | 0.000970 |

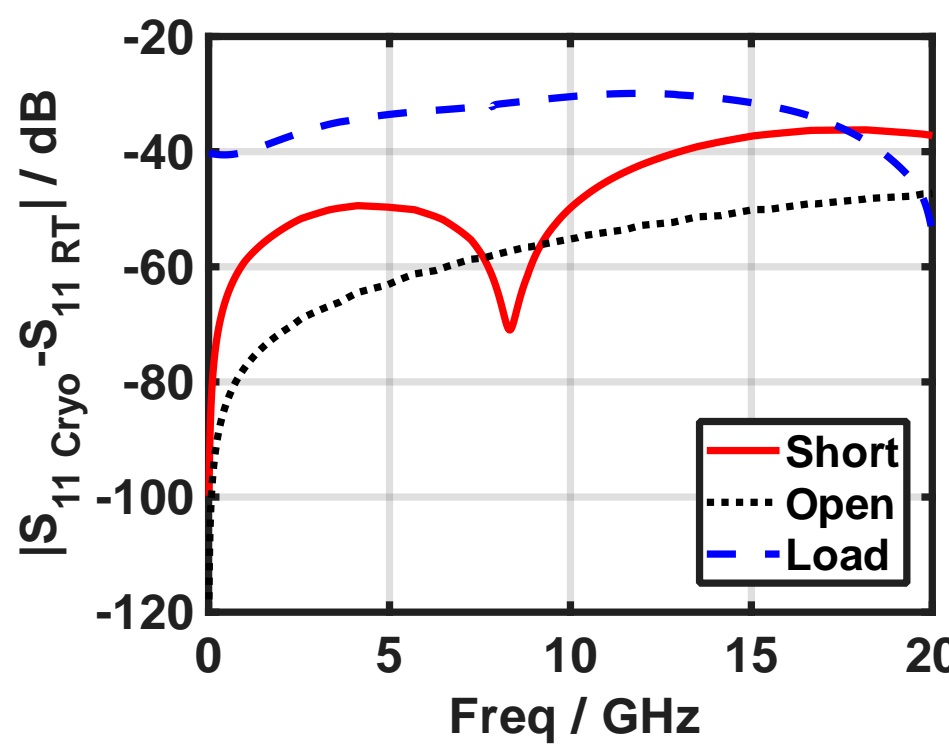

**Fig. 5.** Nominal response shift computed as $|S_{11cryo}-S_{11RT}|$ dB obtained by the EM simulations of the 3D models of the various calibration artifacts.

It is important to mention that when the RT measurements and data employed in the simulation environment (i.e., mechanical dimensions, and the material parameters) are obtained from measurements or other sources maintaining SI traceability, the resulting model response is a best effort to preserve that traceability. Moreover, the shift in response from RT to cryo can be employed as an uncertainty expansion that preserves a link to traceable data. Following this approach, we consider the resulting measurements to retain at least partial traceability to the SI.

## IV. VNA CALIBRATION AND EVALUATION OF THE MEASUREMENT UNCERTAINTY

The system, comprising the VNA and the cryogenic setup, can be regarded as another, yet more complex, VNA. The model describing the instrument, and the calibration process used in this work is described in detail in [20]. It is based on an $N$-port model in which uncertainties are assigned to the error terms, and is described by the following equation:

$$\boldsymbol{M}^{(i)} = \boldsymbol{R}^{(i)} + \left[\left(\boldsymbol{W} + \boldsymbol{V}^{(i)}\right) \oplus \left[\left(\boldsymbol{E} + \boldsymbol{D}^{(i)}\right) \oplus \left[\boldsymbol{C}^{(i)} \oplus \boldsymbol{S}^{(i)}\right]\right]\right], \quad (3)$$

in which $i$ is the measurement index, $\boldsymbol{M}$ contains the raw data measured by the VNA, $\boldsymbol{R}$ denotes the noise and linearity influences, $\boldsymbol{W}$ denotes the switch terms, $\boldsymbol{V}$ describes the drift of the switch terms, $\boldsymbol{E}$ contains the calibration error terms, $\boldsymbol{D}$ describes the drift of the calibration error terms, $\boldsymbol{C}$ denotes the cable stability and connector repeatability, and $\boldsymbol{S}$ denotes the error corrected data or the calibration kit standard definitions. The operator $\oplus$ denotes the cascading of two S-parameter sets.

The system of equations needed for SOLR calibration [6], can be set up starting from (3) and the inverse function of (3) can be used for the error correction:

$$\boldsymbol{S}^{(i)} = \left[\left[\left(\boldsymbol{M}^{(i)} - \boldsymbol{R}^{(i)}\right) \ominus \left(\boldsymbol{W} + \boldsymbol{V}^{(i)}\right)\right] \ominus \left(\boldsymbol{E} + \boldsymbol{D}^{(i)}\right)\right] \ominus \boldsymbol{C}^{(i)}, \quad (4)$$

in which the operator $\ominus$ denotes the de-cascading of two S-parameter sets.

In the measurement model, uncertainty influences are described by $\boldsymbol{R}$, $\boldsymbol{V}$, $\boldsymbol{D}$, $\boldsymbol{C}$, and $\boldsymbol{S}$ and are linearly propagated through calibration and error correction [20].

For such calculations, instead of developing specific software, we use VNATools, a program developed by the Swiss National Metrology Institute METAS [21] – [22]. VNATools is fully compliant with both the EURAMET cg-12 Guide [23] and the Guide to the Expression of Uncertainty in Measurement (GUM), along with its supplements [24].

The uncertainty contributions are evaluated in the following sections, similarly to RT VNAs. Specifically, we follow the widely accepted EURAMET Calibration Guide no. 12 (cg-12) [23]. Details on evaluating various uncertainty contributions are provided in the Guide and are not reiterated here. However, there are differences compared to standard laboratory practices at RT. For instance, there are no test port cables, and the role of connector repeatability is replaced by the contribution from the cryogenic switches.

The uncertainty contributions considered here include Calibration Standards, Noise, Drift, Linearity, and Switches. For Noise, Drift and Linearity contributions, a degradation in performance compared to RT measurements is observed, as expected, due to the presence of the HEMT and LNA amplifiers in the circuit. Fig. 6 to Fig. 9 show a comparison of the RT (VNA only) and mK (VNA and cryogenic system) contributions, evaluated according to EURAMET cg-12.

The sole exceptions are the calibration standards (already discussed in Section III.B), and the linearity (Fig. 10), which was evaluated using the method described in Section IV.C.

In Section V, while discussing the measurement results of an actual test device, the sources of uncertainty with the greatest impact will be identified.

### *A. Noise*

Fig. 6 and 7 show RT (orange) and mK (blue) contribution of Noise Floor and Trace Noise. Noise floor denotes random fluctuations in the absence of a deterministic signal. Trace noise denotes random fluctuations of the measurement result. These fluctuations include short-term temperature variations. They are evaluated according to the procedure in [23, Annex G.1] by short-circuiting the system measurement ports, performing repeated measurements of the four S-parameters, and calculating the standard deviations at each frequency. The worsening with respect to RT is clearly visible.

### *B. Drift*

Fig. 8 and 9 show RT (orange) and mK (blue) contribution of Drift in the reflection (Directivity) and in the transmission (Tracking) coefficients. Drift effects are dependent on the time elapsed after the calibration of the VNA and represent the drift of the measurement results due to the VNA itself, environmental factors, measurement setup and operating conditions. The effect of temperature drift is, therefore, included. The Drift contribution is evaluated according to [23, Annex G.3] over a 24-hour period by measuring the full set of S-parameters of a thru connection, after a two-port calibration, every 15 minutes. For each frequency, the drift of both reflection and transmission parameters is calculated as the maximum deviation from the first measurement ($t = 0$). Again, the worsening with respect to RT is clearly visible.

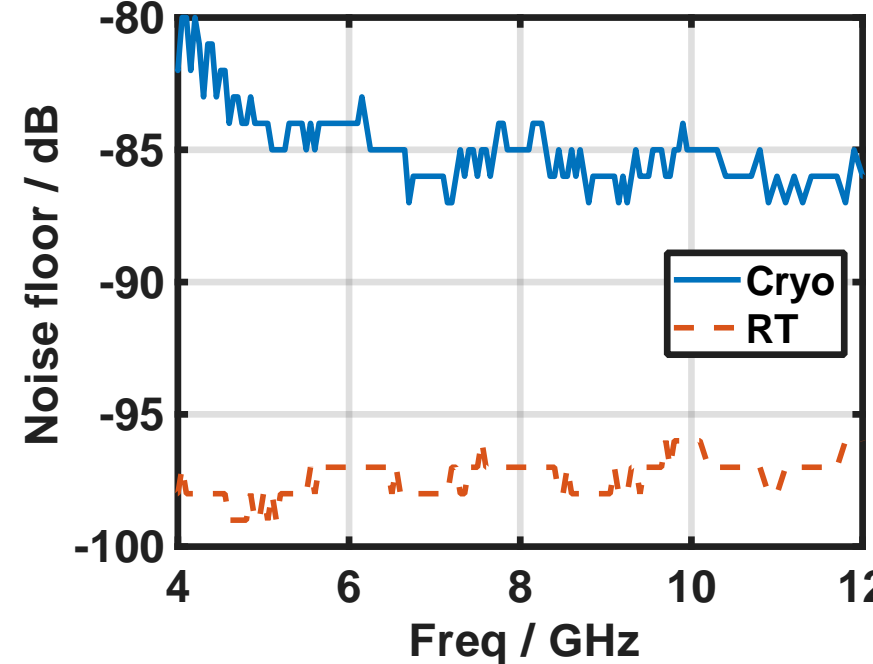

**Fig. 6.** Noise Floor of the VNA only (at RT, orange dashed curve) and of the whole cryogenic system (at mK, blue curve).

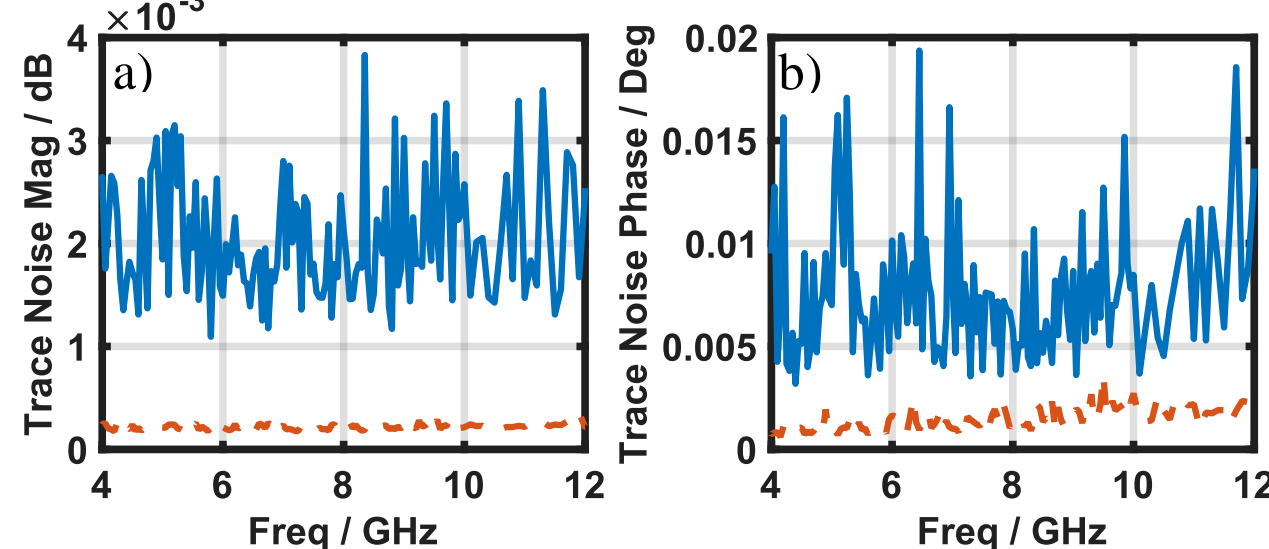

**Fig. 7.** Contribution of the Trace Noise to the S-parameters magnitude (a) and phase (b). In orange (dashed) the VNA only (at RT) and in blue the whole cryogenic system (at mK).

### *C. Linearity*

The measurement system comprising the VNA and the cryogenic setup is modeled as a linear network. The linearity contribution denotes deviations from this behavior. Typically, the linearity error of a VNA is evaluated using a calibrated step attenuator, and it is considered frequency-independent [23, Annex G2]. Unfortunately, this method is not feasible within a cryostat.

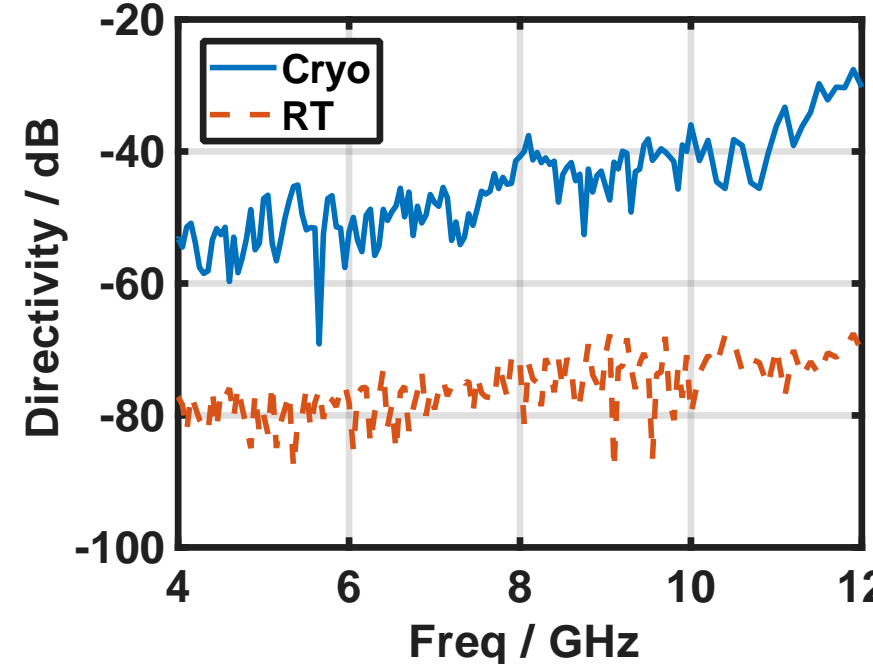

**Fig. 8.** Directivity of the VNA only (at RT, orange dashed curve) and of the whole cryogenic system (at mK, blue curve).

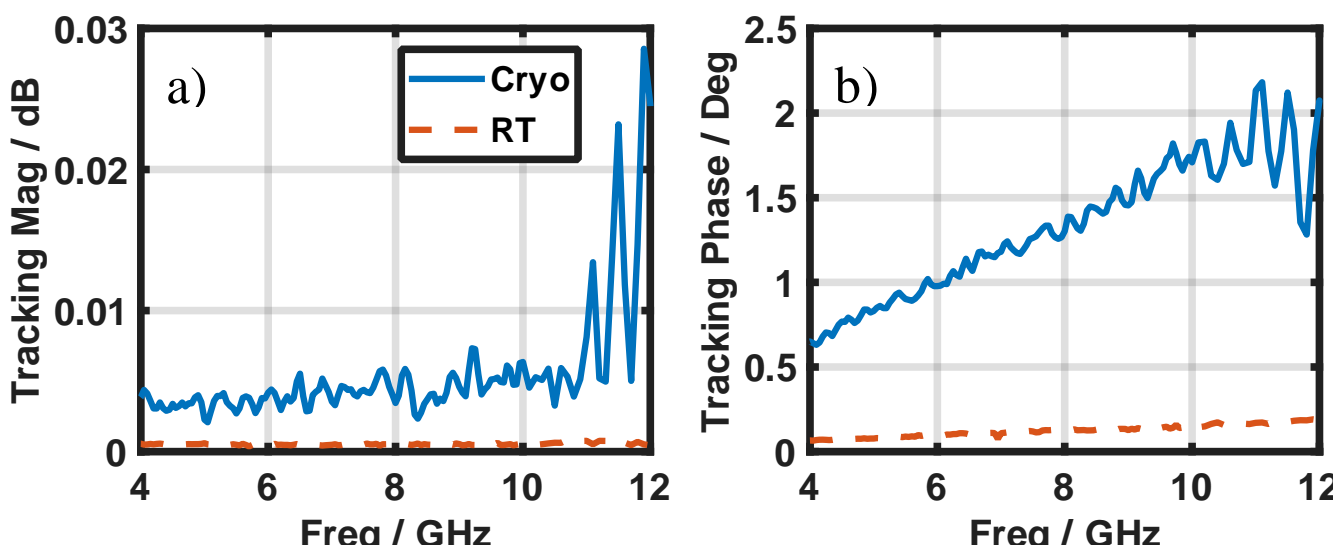

**Fig. 9.** Contribution of the Tracking to the S-parameters magnitude (a) and phase (b). In orange (dashed) the VNA only (at RT) and in blue the whole cryogenic system (at mK).

Anyway, modern VNAs are highly linear, and their linearity error is likely negligible compared to the error introduced by the HEMTs and LNAs in our system. Therefore, we opted to vary the output power of the VNA, utilizing its reference receivers to measure $A_1$ and $A_2$ signals (see Fig. 1), and measured the output signals $B_1$ and $B_2$ by means of the corresponding receivers of the two VNA ports. If the system is linear, the corresponding S-parameter should remain consistent across all power levels.

Fig. 10 illustrates the linearity contribution as a function of both frequency and power. As shown, it is no longer frequency independent.

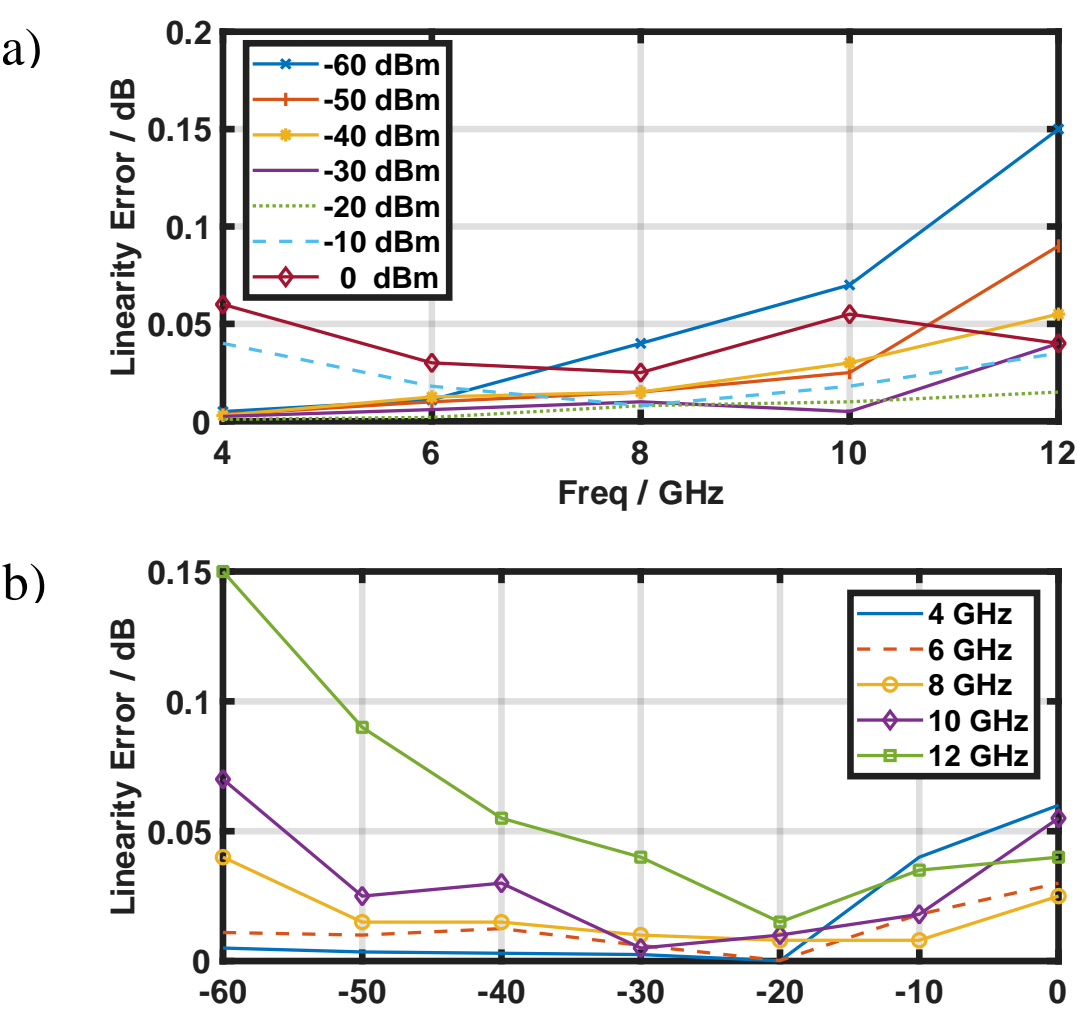

**Fig. 10.** Linearity error with respect to frequency at different power levels (a) and with respect to power at different frequencies (b).

*D. Switches*

As previously mentioned, opening and closing the ports of an RF coaxial switch serves the same function as the repeatability of connectors when using a coaxial cable in standard practice. Additionally, since the paths through the switch ports can differ, the contribution of the switches also affects transmission measurements, with any imbalance between the switch ports having an impact. The effect of switches on measurement results was pointed out in [2]. It was, then, evaluated in [25] in a system in which jumper cables are used to connect calibration standards and DUT to the switches. This study evidenced the emergence of oscillations in both one-port and two-port measurements due to imperfect definition of the calibration plane and proposed an effective mitigation method. Nevertheless, our system is equipped with Radiall Cryogenic SP6T switches (Part Number R5927B2141), and calibration standards are directly connected to the switch ports, without any jumper cables. We did not observe such oscillation, indicating that they are mainly due to differences among the jumper cables.

To assess the contribution of the asymmetry in the RF responses of different switch positions in our setup, repeated measurements of the switch ports S-parameters were conducted at RT and evaluated in [19]. Because the switches are permanently integrated into the cryostat probe and the available space does not allow a conventional two-port characterization, their reflection and transmission responses were evaluated at RT by means of a one-port, two-tier VNA calibration procedure with repeated measurements, in order to estimate repeatability and port-to-port variability. An equivalent repeated characterization at cryogenic temperature was not feasible. Moreover, since each switch actuation perturbs the thermal equilibrium and requires approximately 15 minutes for the system to return to the operating temperature, this makes a statistically meaningful repeated characterization of the six switch ports at cryogenic temperature impractical.

Fig. 11 illustrates the impact of switch port repeatability and port path differences on $S_{11}$ and $S_{21}$ measurements. Fig. 11a shows the standard deviation of the magnitude of $S_{11}$ for 30 measurements for each channel (switch ports), while Fig. 11b shows $S_{21}$ of each switch channel and the corresponding maximum difference (right axis). Data for port 5 are not reported, as this switch port was non-functional due to a hardware issue. However, since only five switch ports were required for these measurements, port 5 was not utilized.

Although, according to the manufacturer, temperature variations are expected to have a minor impact, the room-temperature (RT) uncertainties still represent the best achievable estimate given current measurement capabilities.

Crosstalk may also be present among ports, but this contribution was not evaluated.

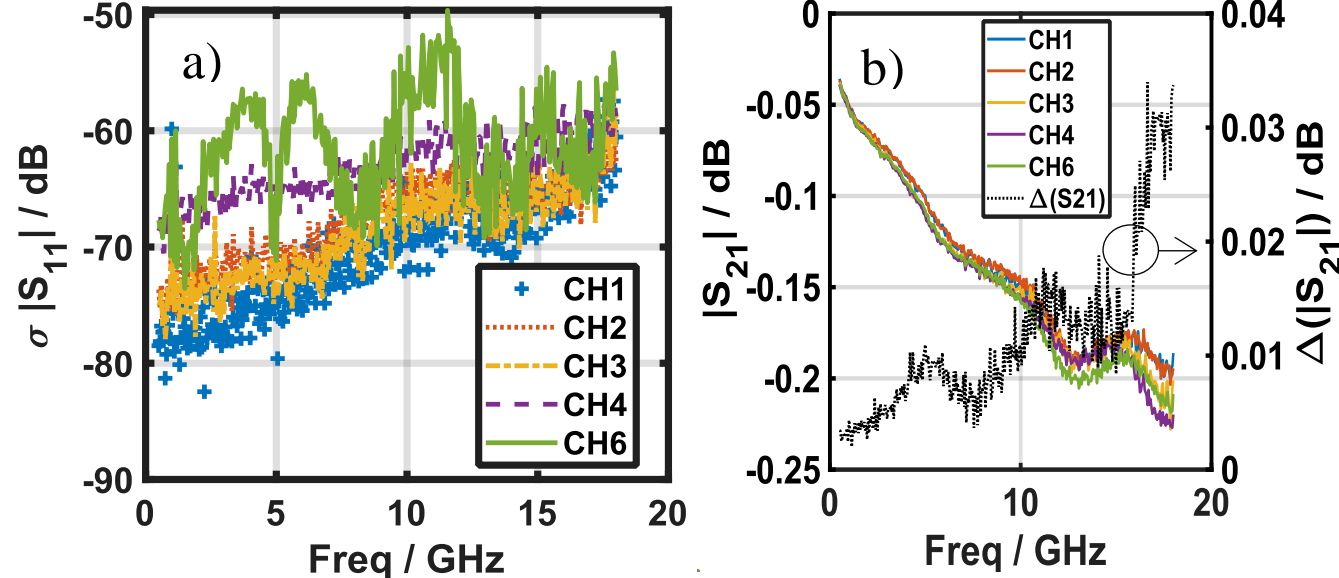

**Fig. 11.** Contribution of switch-port repeatability and port-path differences to $S_{11}$ and $S_{21}$.

## V. Measurement Results

Once the calibration standards for cryogenic operation have been defined based on SI-traceable room-temperature measurements, with an additional uncertainty contribution accounting for the response shift upon cooling, and once the relevant uncertainty contributions have been evaluated, the measured data can be corrected through the calibration procedure and associated with an uncertainty budget.

In this section, results of an actual DUT are presented. The DUT is a 20 dB attenuator (Shenzhen Torvenics Electronic Co., Ltd SMA FIXED ATTENUATOR 2 W 18 GHz 20 dB Nickel free P/N: 8221-2W-18G-20(dB)). Measurements were carried out in the full frequency band allowed by the setup, that is from 4 to 12 GHz. As an example, Fig. 12 presents raw (blue curve) and calibrated (orange curve) data for the modulus and phase of $S_{21}$. The measurement uncertainty for the calibrated data is not shown in the figure, as the uncertainty bounds are not discernible due to the graph's scale. Nevertheless, Table II provides the uncertainty budget for $|S_{21}|$ at a frequency of 6 GHz as an example. At this frequency, the measured value is $|S_{21}|$ = -20.70 ± 0.08 dB (95% confidence interval). Figure 13 reports on the behavior of the relative uncertainties over the full frequency range. While the uncertainty at 6 GHz is dominated by calibration standards, switches, and linearity, which contribute almost equally, and an additional significant contribution comes from noise, at lower frequencies the most relevant contribution is linearity.

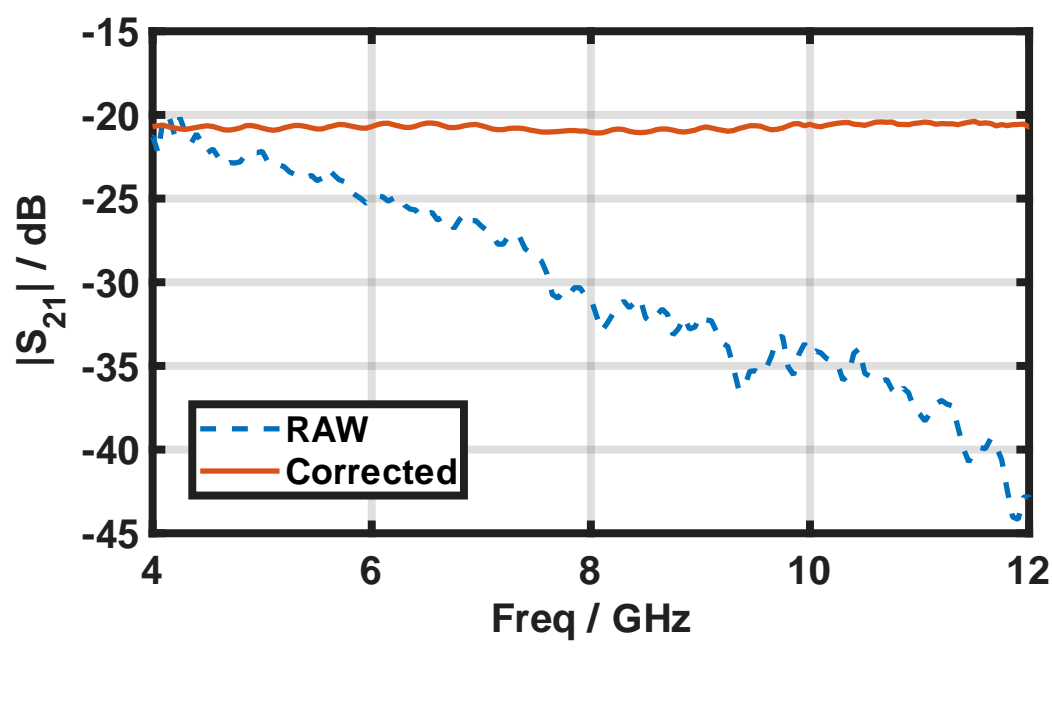

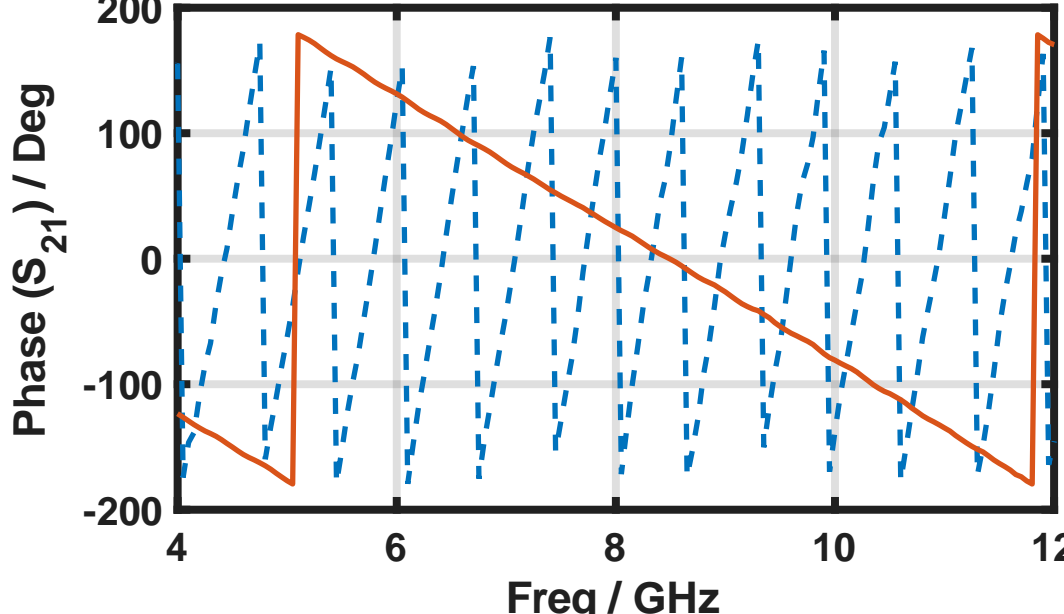

**Fig. 12.** Calibrated (orange curve) and uncalibrated (blue curve) measurement of $S_{21}$ of a 20 dB attenuator used as DUT.

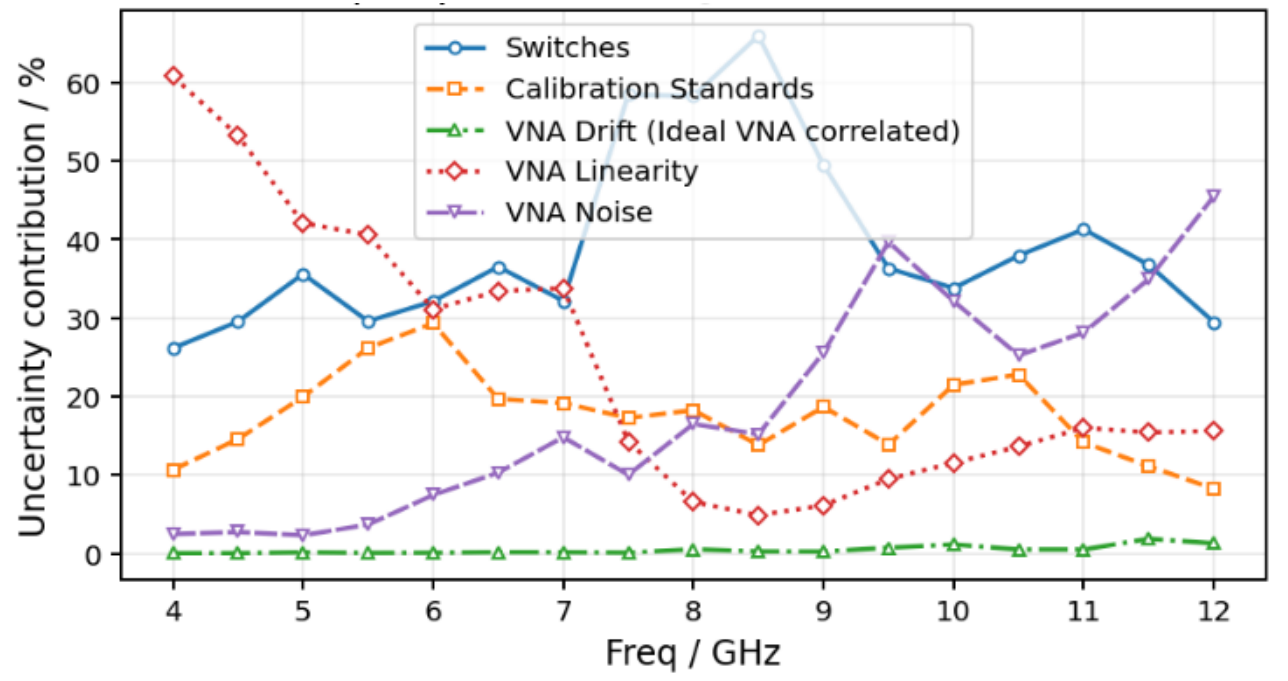

**Fig. 13.** Relative uncertainty contributions to $|S_{21}|$ over the full frequency range.

TABLE II

BREAKDOWN OF THE UNCERTAINTY SOURCES FOR THE MEASUREMENT OF $|S_{21}|$ OF THE DUT AT 6 GHZ

| Uncertainty source | Uncertainty Contribution / dB | Uncertainty Percentage |
|---|---|---|
| Calibration Standards | 0.021699 | 29.306 |
| Switches | 0.022710 | 32.100 |
| Drift | 0.001149 | 0.082 |
| Linearity | 0.022347 | 31.083 |
| Noise | 0.010925 | 7.430 |

TABLE III

CONTRIBUTION OF THE CALIBRATION STANDARDS TO THE MEASUREMENT OF $|S_{21}|$ OF THE DUT AT 6 GHZ

| Uncertainty source | Uncertainty Contribution / dB | Uncertainty Percentage |
|---|---|---|
| Short | 0.011143 | 7.729 |
| Open | 0.011441 | 8.148 |
| Load | 0.014689 | 13.429 |

Conversely, at high frequencies system noise appears to have also a major role. Table III highlights the contributions from the three types of calibration standards. As illustrated, the Load standards give the most relevant contribution, while Shorts and Opens contribute almost equally to the overall uncertainty.

Moreover, Table IV presents the uncertainty budget for $|S_{11}|$ of the same DUT at the same frequency of 6 GHz. The measured value is $|S_{11}|$ = -11.98 ± 0.87 dB (95% confidence interval). In Fig. 14 the behavior of the relative uncertainties over the full frequency range is shown. In this case, the most significant contribution to the uncertainty comes from the switches. Calibration standards also have a significant impact. Additionally, an examination of calibration standards' contribution (see Table V) reveals that the loads play again a major role.

This budget analysis enables the identification of the most critical aspects for improvement in future work: the definition of the Load calibration standard, the contribution of the switches, and the evaluation of linearity. The noise floor and

TABLE IV

BREAKDOWN OF THE UNCERTAINTY SOURCES FOR THE MEASUREMENT OF $|S_{11}|$ OF THE DUT AT 6 GHZ

| Uncertainty source | Uncertainty Contribution / dB | Uncertainty Percentage |
|---|---|---|
| Calibration Standards | 0.179875 | 17.276 |
| Switches | 0.390458 | 81.402 |
| Drift | 0.013896 | 0.103 |
| Linearity | 0.047015 | 1.180 |
| Noise | 0.008542 | 0.039 |

TABLE V

CONTRIBUTION OF THE CALIBRATION STANDARDS TO THE MEASUREMENT OF $|S_{11}|$ OF THE DUT AT 6 GHZ

| Uncertainty source | Uncertainty Contribution / dB | Uncertainty Percentage |
|---|---|---|
| Short | 0.078731 | 3.310 |
| Open | 0.065756 | 2.309 |
| Load | 0.147759 | 11.657 |

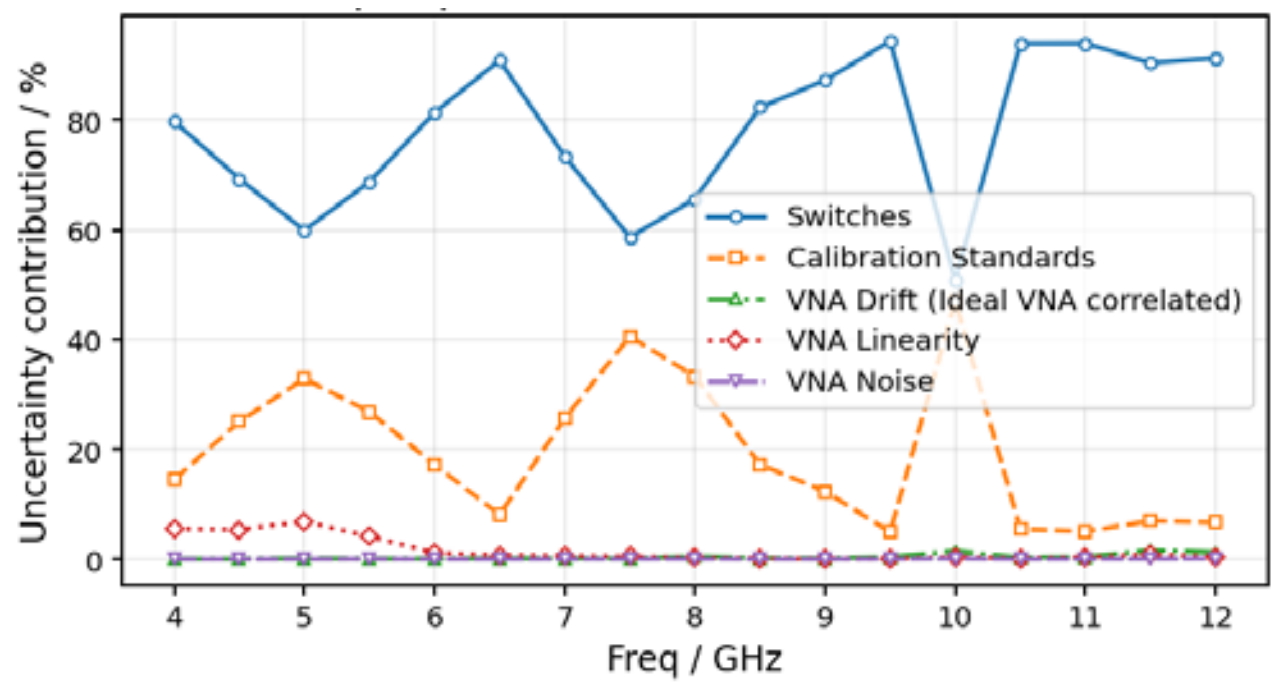

**Fig. 14.** Relative uncertainty contributions to $|S_{11}|$ over the full frequency range.

trace noise may also have a significant impact, while system drift remains generally negligible. However, performing the measurements takes a long time due to the necessary waiting period needed for thermalization after each switch movement, which lasts approximately 15 minutes.

A comparison of $|S_{21}|$ and $|S_{11}|$ at 45 mK and at RT (see Table VI and Fig. 15 and 16) shows that the attenuation at mK is 0.3-1 dB higher (i.e., about 1.5-5%). At the considered frequency of 6 GHz, the difference is 0.67 dB (i.e., about 3.4%). Considering that the relative expanded uncertainty of $|S_{21}|$ at 6 GHz is 0.4%, as discussed above, this difference is significant. From Fig. 16, a clear worsening of the attenuator matching condition is also noticeable. These results indicate that the attenuator DUT undergoes a significant change when cooled to mK temperatures. This behavior is plausibly explained by the fact that the DUT is a fixed attenuator based on a resistive network: non-uniform cryogenic changes in the resistor values can modify both the nominal attenuation ratios and the matching conditions of the device, thereby affecting both $S_{11}$ and $S_{21}$.

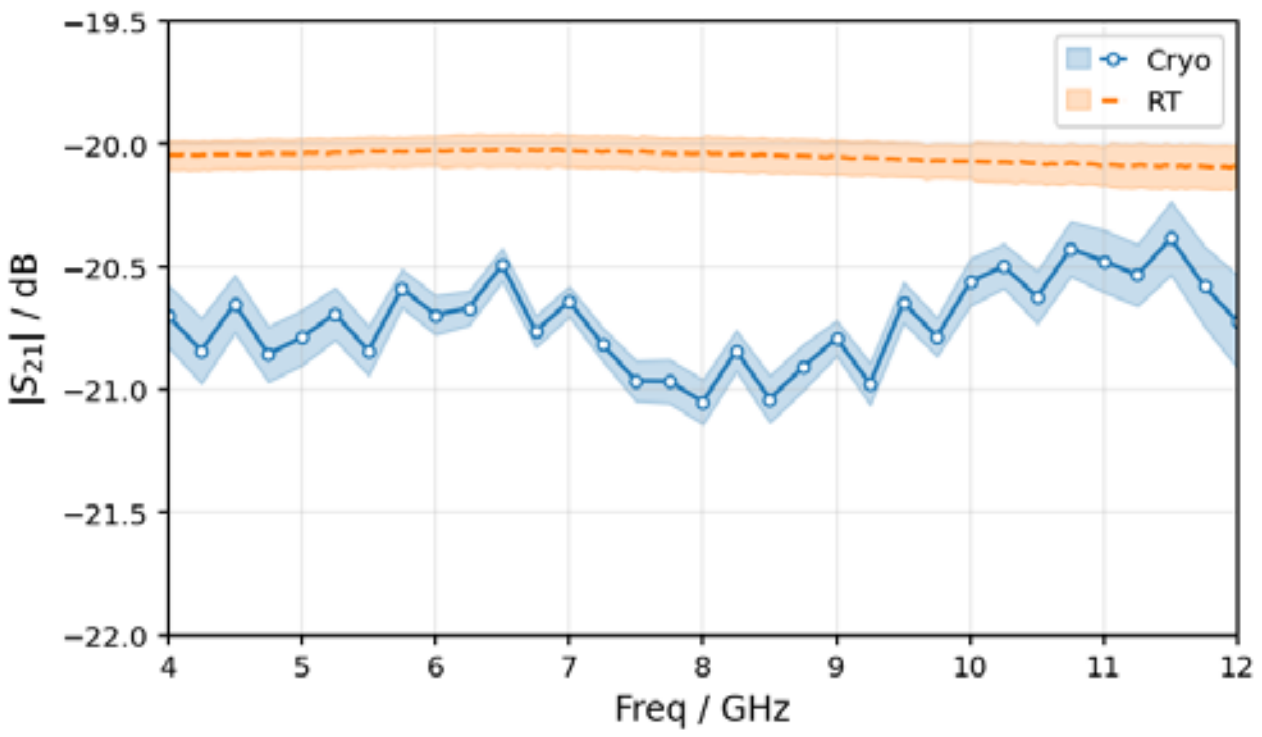

**Fig. 15.** Room temperature and cryogenic measurements of $|S_{21}|$ of the DUT across the full frequency range.

TABLE VI

ROOM-TEMPERATURE AND CRYOGENIC MEASUREMENTS OF $|S_{21}|$ OF THE DUT AND RELATED EXPANDED UNCERTAINTIES

| Frequency / GHz | $\|S_{21}\|_{RT}$ / dB | $\|S_{21}\|_{CRYO}$ / dB |
|---|---|---|
| 4 | -20.045 ± 0.06 | -20.699 ± 0.13 |
| 6 | -20.026 ± 0.06 | -20.696 ± 0.08 |
| 8 | -20.034 ± 0.07 | -21.050 ± 0.09 |
| 10 | -20.068 ± 0.07 | -20.561 ± 0.10 |
| 12 | -20.089 ± 0.09 | -20.724 ± 0.19 |

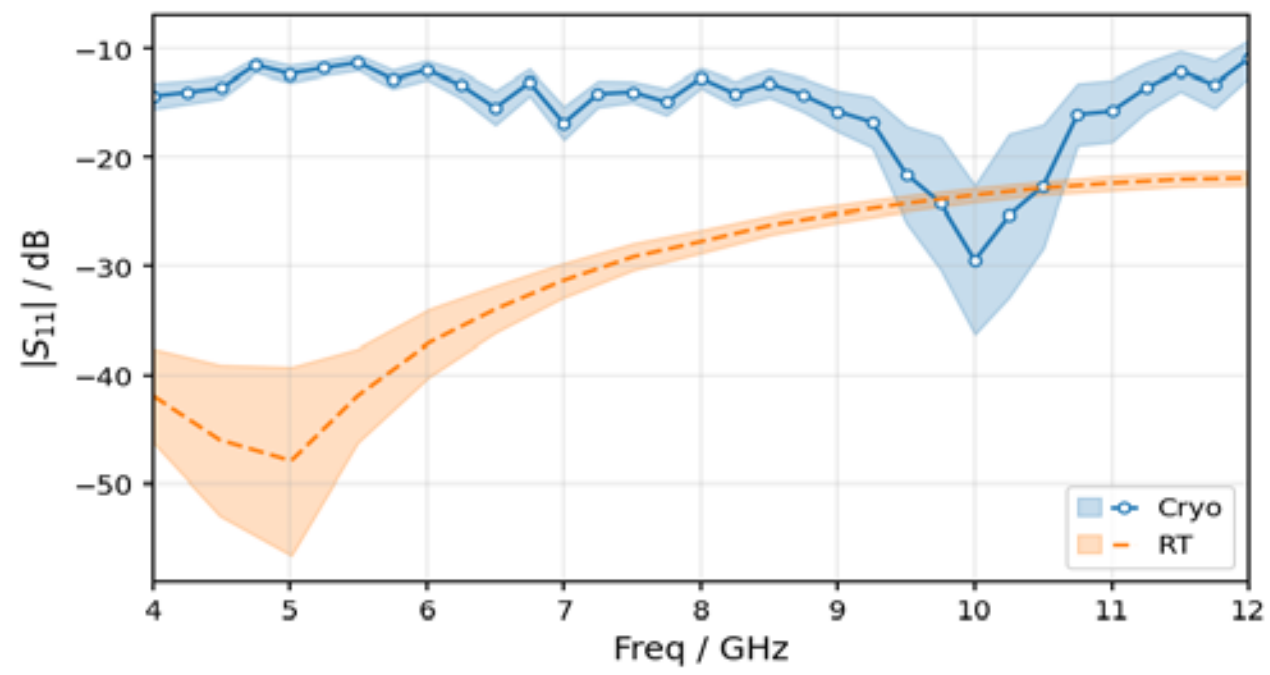

**Fig. 16.** Room temperature and cryogenic measurements of $|S_{11}|$ of the DUT across the full frequency range.

It should be noted that the RT measurements of the DUT reported in Table VI, Fig. 15 and 16 were performed in a standard bench-top VNA configuration rather than with the full cryogenic measurement chain, since the HEMT amplifiers used in the cryogenic setup cannot be operated at RT under the same working conditions.

The residual oscillations visible in Fig. 15 and 16 in the cryogenic measurements are interpreted as arising from mismatch in the jumper cables connecting the DUT to the switch ports. These are very thin and fragile sections of CuNi/CuNi coaxial cable about 12 cm long and with an outer diameter of 1.19 mm, and are the same kind of cables used as Unknown Thru standards. It is likely that their repeated bending during assembly and disassembly altered their impedance, thereby producing standing waves. Separate measurements of the cables confirmed the interpretation. These oscillations are not visible in the coincidence-test one-port measurements where jumper cables are not used (see Section VI). Since this effect is attributed to jumper-cables failure that will be eliminated in the future by choosing different cables, this contribution is not included in the uncertainty budgets to highlight the other influencing factors.

A practical question is whether it is really necessary to use as many as four switch ports to perform a SOLR calibration, rather than using just one to apply a Thru-based complex correction and using the remaining free ports to measure multiple devices in a single cooldown. A comparison between complex Thru correction and SOLR calibration for the 20 dB attenuator measurements shows that the discrepancy between the two approaches remains within approximately ±0.6 dB across the entire frequency range (see Fig. 17). In view of the considerations discussed in the previous paragraphs, such a discrepancy cannot be regarded as negligible in the present measurement context. Moreover, although complex Thru correction provides magnitude and phase information for the transmission response, it does not correct source/load-match effects associated with DUT mismatch and remains limited to transmission measurements only. Therefore, it cannot provide the complete S-parameter characterization ensured by SOLR calibration. Phase-resolved characterization is often important for the extraction of circuit parameters of DUTs, as illustrated, for example, in [26] – [27].

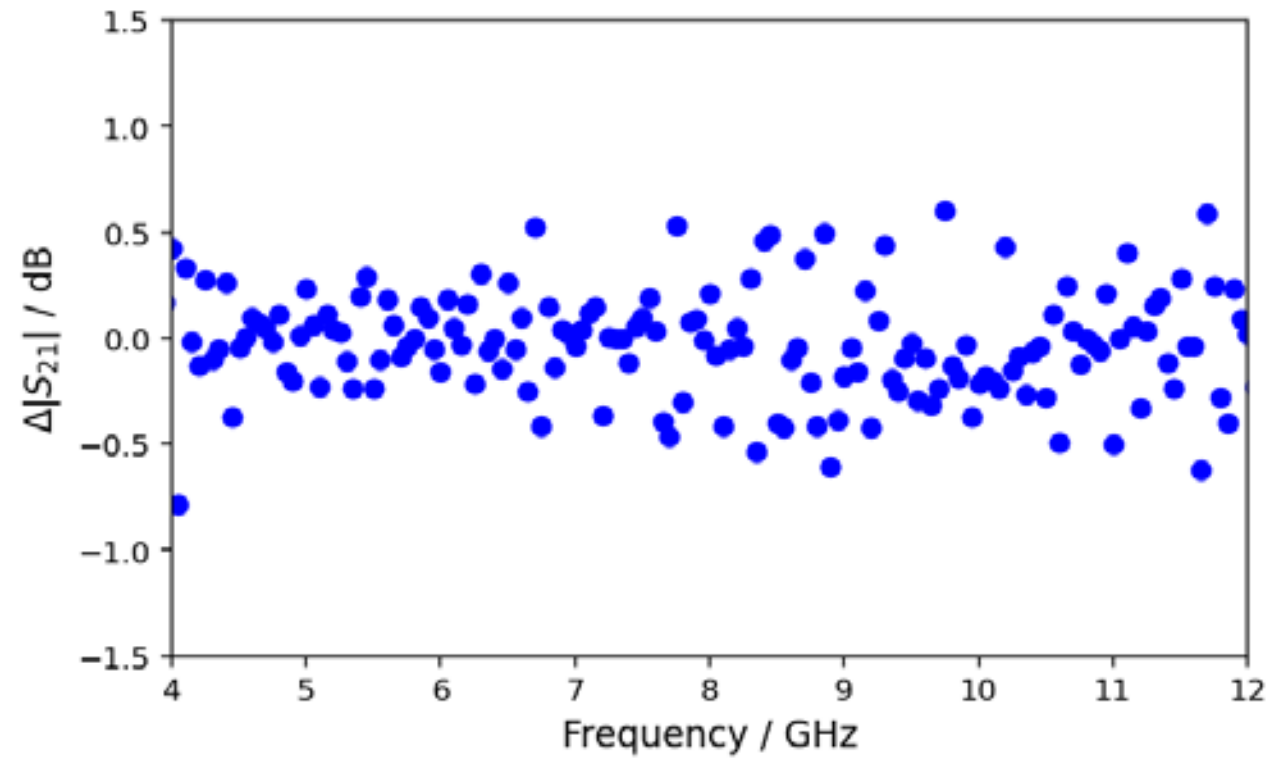

**Fig. 17.** Difference $\Delta|S_{21}| = |S_{21}|_{\text{Thru}} - |S_{21}|_{\text{SOLR}}$, expressed in dB, between the $|S_{21}|$ values of the 20 dB attenuator DUT obtained at cryogenic temperature by Thru normalization and by SOLR calibration.

## VI. Towards Calibration Verification

According to [23], VNA calibration can be verified by means of coincidence tests and plausibility tests. In coincidence tests, a set of SI-traceable verification standards with known reference values, such as attenuators, adapters, mismatched terminations, and offset shorts, is typically employed to enable a quantitative assessment by comparing the measured responses with the reference values. By contrast, a plausibility test provides a qualitative assessment of the validity of the VNA calibration based on fundamental physical considerations.

A viable approach to verification under cryogenic conditions would be to use an airline, whose SI traceability derives from its mechanical dimensions. The use of airlines at cryogenic temperatures as TRL calibration devices was already demonstrated in [28], while their suitability as primary impedance standards at mK temperatures was investigated in [29]. In [19] a numerical approach was used to evaluate the behavior of an airline at 77 K and 4.2 K, together with the associated S-parameters measurement uncertainty; however, the method was not tested at mK temperatures. Another possible plausibility strategy is to compare results obtained with two independent cryogenic calibration techniques, as done in [30], where mK S-parameter measurements were validated by agreement between independent TRL-based approaches. Such an approach would still provide an initial validation rather than a full SI-traceable verification, but it could increase confidence in the measurement results. In the present setup, however, both airline-based verification and alternative independent calibration routes would require connection through jumper cables. Since these cables cannot be assumed to be perfectly identical, their mismatch and path-to-path variability would introduce additional errors, thereby limiting the effectiveness of such verification artifacts or comparison procedures in the present configuration.

Given the current lack of established two-port standards for VNA verification at mK temperatures, we leveraged the findings reported in Section III.C and performed both a coincidence test and a plausibility test. It was found that the Short and Open standards we used for calibration undergo negligible changes when cooled from RT to mK temperatures. Therefore, this kind of standards can be used for an initial one-port coincidence test against known reference data. Measurement of a Flush Short is suitable for checking whether the reference plane has been correctly established [23], since $|S_{11}|$ is expected not to exceed unity and its phase should ideally be equal to 180°. The Open device, which introduces a shunt capacitance due to its mechanical characteristics [31], exhibits a frequency-dependent phase response that can also be checked in a coincidence test.

Two standards, an Open and a Short, of the same type as those used in Section III.C and from the same manufacturer, were measured. For this specific measurement, calibration standards were connected to switch ports 3, 4 and 6 (Open, Load and Short, respectively) while the verification devices were connected to switch port 1 (Open on the VNA port 1 side, Short on the VNA port 2 side). Fig. 18 a) and b) compare their magnitude and phase measured at RT (red curve, SI-traceable) and at mK (blue curve). Very good agreement is observed for both standards in both magnitude and phase. The absence of residual oscillations confirms that no visible effects arise from switch-port asymmetries.

For two-port measurements, the calibration can be qualitatively assessed by exploiting the reciprocity of our DUT, which provides a means of evaluating possible asymmetries in VNA calibration. For a reciprocal DUT, $S_{21} = S_{12}$ is expected; any significant deviation would point to residual asymmetries or errors in the calibration. The results of this plausibility test are shown in Fig. 19. The observed deviations are negligible, remaining below 0.0025 in linear dimensionless units. Although this reciprocity-based comparison does not constitute an independent uncertainty evaluation of the transmission measurement, it provides a qualitative check of the internal consistency of the calibrated two-port response. In particular, the small observed deviation between $S_{21}$ and $S_{12}$ indicates that residual calibration asymmetries are limited and is consistent with the uncertainty budget reported for the DUT transmission measurement. Therefore, Fig. 19 supports the validity of the calibrated DUT transmission measurements.

However, full verification of S-parameter measurements in coaxial lines at cryogenic temperatures to ensure SI traceability remains an open challenge.

## VII. Conclusion

The presented work has detailed the development and characterization of a cryogenic millikelvin coaxial S-parameters measurement system. The system operates in the 4-12 GHz range and was developed and installed at INRiM in a collaborative effort with Delft University of Technology.

The developed system leverages direct access to VNA receivers, low-cost commercially available calibration standards, and electromechanical switching. The databased definition of calibration standards, exploiting SI-traceable RT measurements, and electromagnetic simulations, allows the study of the calibration standards behavior when the temperature is drastically reduced from RT to mK. If the data employed in the simulation environment are obtained from measurements or other sources maintaining SI traceability, the resulting model response is a best effort to preserve that traceability. Moreover, the shift in response from RT to mK can be employed as an uncertainty expansion that preserves a link to traceable data. This approach allows the use of internal VNA calibration routines or existing software (e.g., METAS VNATools) avoiding the need for dedicated software.

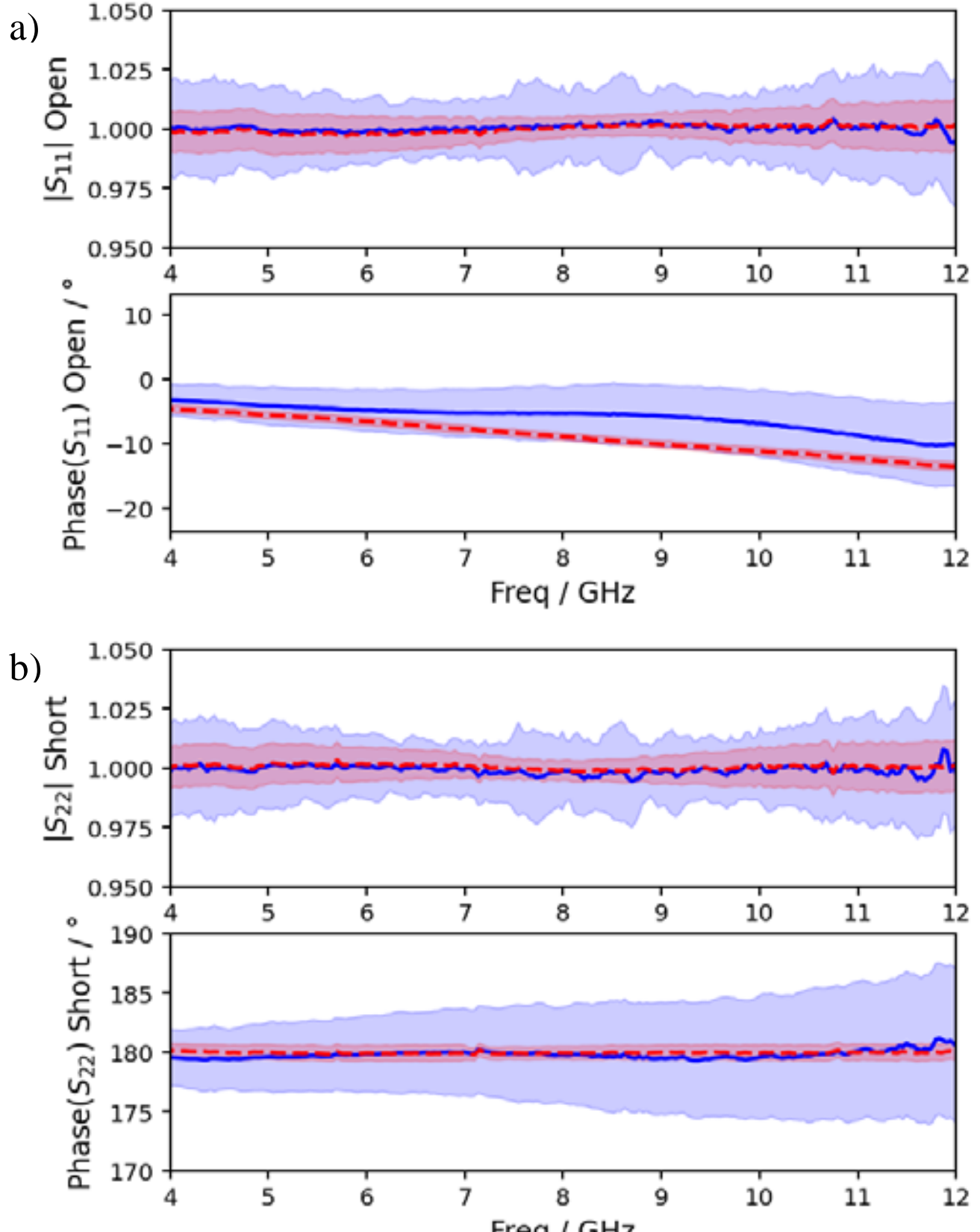

**Fig. 18.** Comparison between RT (red curve) and mK (blue curve) measurements of a) XMA cryogenic Open device, and b) XMA cryogenic Short device.

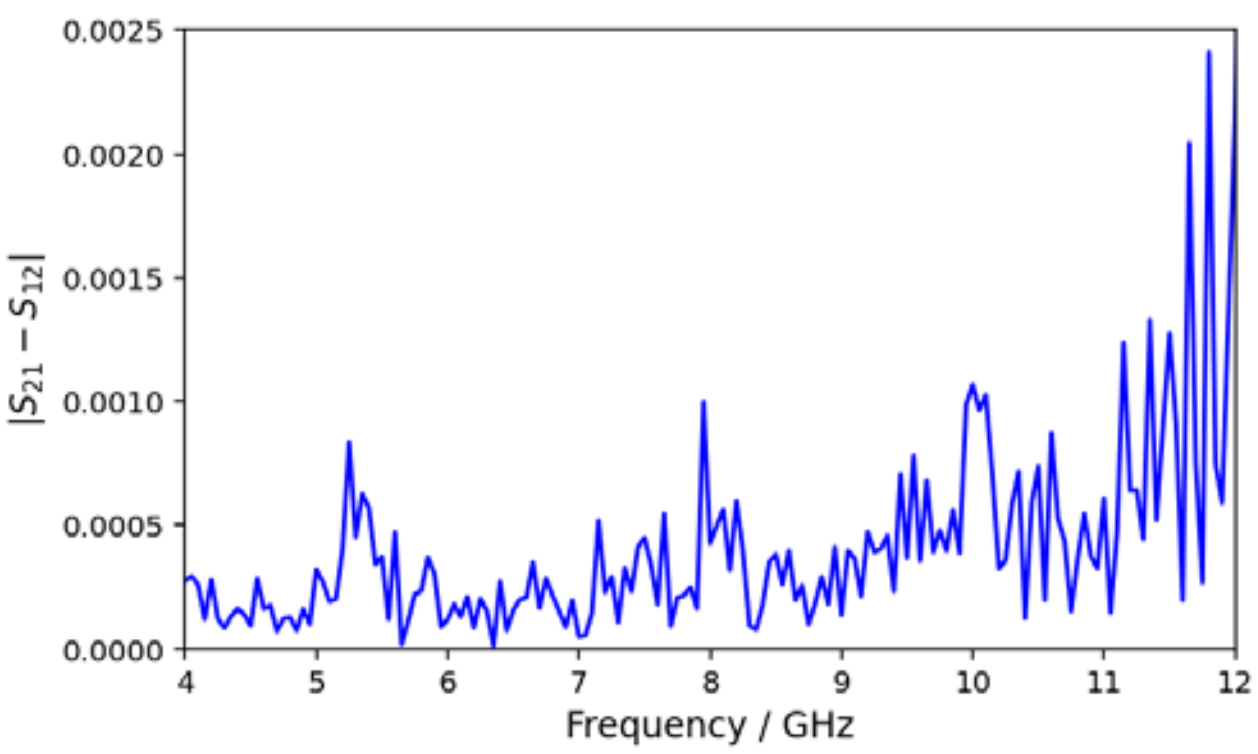

**Fig. 19.** Reciprocity-based plausibility test for the calibrated transmission response of the DUT. Comparison between the measured forward and reverse transmission coefficients shows deviations below 0.0025 in linear dimensionless units.

The presented results show that a full two-port SOLR calibration procedure can be implemented at cryogenic temperatures and applied to obtain calibrated two-port S-parameter measurements with an associated uncertainty budget. A comprehensive uncertainty budget was evaluated and detailed, including the results of the evaluation of different uncertainty components, for the first time to the best of the authors' knowledge. Additionally, calibrated measurements of a 20 dB attenuator used as test DUT were provided, demonstrating the system's capability to assess the measurement uncertainty of all four S-parameters and to identify the most significant contributors to the uncertainty. Finally, a comparison was made between RT and cryogenic measurements, as well as between cryogenic SOLR calibration and complex Thru normalization demonstrating the advantage of the proposed methodology.

While full verification of S-parameter measurements in coaxial lines at cryogenic temperatures to ensure SI traceability remains an open challenge, an initial verification of the VNA calibration was carried out by combining a quantitative coincidence test against known SI-traceable standards for one-port measurements with a qualitative plausibility test of the transmission response. Some limitations of the present work should also be emphasized. Unlike [8], the present work does not investigate a wide temperature range, but is specifically focused on the mK regime, which is particularly relevant for the operation and characterization of superconducting quantum devices and microwave circuits [3]. Moreover, the current implementation is limited to two-port measurements, whereas recent developments have addressed multiport cryogenic characterization [9]. In addition, the calibration verification presented here is initial in nature and does not rely on independent cryogenic calibration techniques as in [30]. The use of airline-based verification devices in the present setup would require jumper-cable interconnections, whose mismatch and lack of perfect identity would introduce additional errors. Finally, switch-related effects are currently estimated from room-temperature measurements only, and a direct cryogenic characterization of switch asymmetries remains future work. In this respect, the present work also differs from the explicit traceability-chain framework discussed in [10], since a complete traceability framework for cryogenic coaxial S-parameter measurements has not yet been established here.

## Acknowledgment

L. O., E. E. and L. F. Authors thank L. Ranzani (RTX BBN Technologies) and A. Giachero (University of Milano Bicocca) for useful discussions.

## References

[1] M. Bieler, et al., "Microwave metrology for superconducting quantum circuits," in *Proc. CPEM2022*, Wellington, New Zealand, *2022*, p. 463.

[2] L. Ranzani, L. Spietz, Z. Popovic and J. Aumentado, "Two-port microwave calibration at millikelvin temperatures," *Rev. Sci. Instrum.*, 84, 034704, Mar. 2013, doi: 10.1063/1.4794910.

[3] H. Wang, et al., "Cryogenic single-port calibration for superconducting microwave resonator measurements," *Quantum. Sci. Technol.*, 6, 035015, Jun. 2021, doi: 10.1088/2058-9565/AC070E.

[4] S. Simbierowicz, V. Y. Monarkha, S. Singh, N. Messaoudi, P. Krantz and R. E. Lake, "Microwave calibration of qubit drive line components at millikelvin temperatures," *Appl. Phys. Lett.*, 120, 054004, Feb. 2022, doi: 10.1063/5.0081861.

[5] M. Stanley, S. De Graaf, T. Hönigl-Decrinis, T. Lindström and N. Ridler, "Characterizing scattering parameters of superconducting

quantum integrated circuits at milli-Kelvin temperatures," *IEEE Access*, 10, pp. 43376 – 43386, Apr. 2022, doi: 10.1109/ACCESS.2022.3169787.

[6] A. Ferrero and U. Pisani, "Two-port network analyzer calibration using an unknown 'thru'," *IEEE Microwave and Guided Wave Letters*, vol. 2, no. 12, pp. 505–507, Dec. 1992, doi: 10.1109/75.173410.

[7] S. Simbierowicz et al., "Calibrated transmission and reflection from a multi-qubit microwave package," *Rev. Sci. Instrum.*, vol. 94, no. 5, 054713, 2023, doi: 10.1063/5.0144840.

[8] T. Arakawa, and S. Kon, "Calibrated Two-Port Microwave Measurement up to 26.5 GHz for Wide Temperature Range From 4 to 300 K," *IEEE Trans. Instr. Meas.*, vol. 72, 1009608, 2023, doi: 10.1109/TIM.2023.3315393.

[9] M. Stanley et al., "Characterizing S -Parameters of Microwave Coaxial Devices With up to Four Ports at Temperatures of 3 K and Above for Quantum Computing Applications," *IEEE Trans. Instr. Meas.*, vol. 73, pp. 1–6, 2024, doi: 10.1109/TIM.2024.3369144.

[10] H. Li et al., "Traceable Uncertainty Propagation and Analysis in S-Parameter Measurement Due to Nonideal Calibration Standards and Noise," *IEEE Trans. Microw. Theory Tech.,* vol. 74, no. 4, pp. 3056–3067, 2026, doi: 10.1109/TMTT.2026.3661040

[11] L. Oberto, et al., "Measurement and Calibration Approaches for Two-Port Scattering Parameters at mK Temperatures," in *Proc. CPEM2024*, 25-05168, Denver (CO), USA, 2024, doi: 10.1109/CPEM61406.2024.10646000.

[12] S. Krinner, et al., "Engineering cryogenic setups for 100-qubit scale superconducting circuit systems," *EPJ Quantum Technology.*, vol. 6, 2, May. 2019, doi: 10.1140/EPJQT/S40507-019-0072-0.

[13] A. Rettaroli, et al., "Ultra low noise readout with traveling wave parametric amplifiers: The DARTWARS project," *Nucl. Instrum. Methods Phys. Res. A*, vol. 1046, 167679, Jan. 2023, doi: 10.1016/J.NIMA.2022.167679.

[14] S. Pagano, et al., "Development of Quantum Limited Superconducting Amplifiers for Advanced Detection," *IEEE Trans. Appl. Supercond.*, vol. 32, no. 4, 1500405, Jun. 2022, doi: 10.1109/TASC.2022.3145782.

[15] V. Granata, et al., "Characterization of Traveling-Wave Josephson Parametric Amplifiers at T = 0.3 K," *IEEE Trans. Appl. Supercond.*, vol. 33, no. 1, 0500107, Jan. 2023, doi: 10.1109/TASC.2022.3214656.

[16] XMA Corp, https://www.xmacorp.com/

[17] "Microwave Studio (MWS). (2022). CST—Computer Simulation Technology AG. [Online]. Available: www.cst.com/products/cstmws."

[18] F. A. Mubarak, V. Mascolo, F. Hussain and G. Rietveld, "Calculating S-Parameters and Uncertainties of Coaxial Air-Dielectric Transmission Lines," *IEEE Trans. Instr. Meas.*, vol. 73, pp. 1-11, 2024, Art. no. 8000511, doi: 10.1109/TIM.2023.3338667.

[19] E. Shokrolahzade et al., "Thermo-Mechanical EM Models for Broadband Cryogenic VNA Calibration Including Numerical Uncertainties Down to 4.2 K," *IEEE Trans. Microw. Theory Tech.*, vol. 73, no. 11, pp. 9058-9069, Nov. 2025, doi: 10.1109/TMTT.2025.3584196.

[20] M. Wollensack, J. Hoffmann, J. Ruefenacht and M. Zeier, "VNA Tools II: S-parameter uncertainty calculation," in *Proc. 79th Microwave Measurement Conference (ARFTG),* Montreal, Canada, 2012, doi: 10.1109/ARFTG79.2012.6291183.

[21] https://www.metas.ch/vnatools

[22] M. Zeier, J. Rüfenacht, M. Wollensack, "VNA Tools – a software for metrology and industry", *METinfo*, vol. 27, no. 2, pp. 4-7, 2020.

[23] EURAMET. Calibration Guide No. 12, Version 3.0 (Mar. 2018). *Guidelines on the Evaluation of Vector Network Analyzers (VNA)*. [Online]. Available: https://www.euramet.org/publications-media-centre/calibration-guidelines

[24] BIPM, IEC, IFCC, ILAC, ISO, IUPAC, IUPAP and OIML. JCGM 100:2008 (2008). *Evaluation of measurement data - Guide to the expression of uncertainty in measurement*. [Online]. Doi: 10.59161/JCGM100-2008E.

[25] M. Stanley et al., "A Technique to Improve Accuracy of S-Parameter Measurements of Coaxial Connectorized Devices at Cryogenic Temperatures," *IEEE Trans. Instr. Meas*., vol. 74, 8005612, 2025, doi: 10.1109/TIM.2025.3595257.

[26] A. Ranadive, et al., "Kerr reversal in Josephson meta-material and traveling wave parametric amplification," *Nat. Commun.*, vol. 13, 1737, Apr. 2022, doi: 10.1038/S41467-022-29375-5.

[27] L. Fasolo, et al., "Experimental Characterization of RF-SQUIDs Based Josephson Traveling Wave Parametric Amplifier Exploiting Resonant Phase Matching Scheme," *IEEE Trans. Appl. Supercond.*, vol. 34, no. 3, 1101406, Jan. 2024, doi: 10.1109/TASC.2024.3359163.

[28] S. H. Shin et al., "Broadband Coaxial S-Parameter Measurements for Cryogenic Quantum Technologies," *IEEE Trans. Microw. Theory Tech.*, vol. 72, no. 4, pp. 2193–2201, 2024, doi: 10.1109/TMTT.2023.3322909.

[29] J. Skinner et al., "Characterizing precision coaxial air lines as reference standards for cryogenic S-parameter measurements at milli-kelvin temperatures," in IEEE MTT-S Int. Microw. Symp. Dig., San Diego, CA, USA, Jun. 2023, pp. 1–4, doi: 10.1109/IMS37964.2023.10188171.

[30] M. Stanley et al., "Validating S-parameter measurements of RF integrated circuits at milli-Kelvin temperatures," *Electronics Letters*, vol. 58, no. 16, pp. 614–616, 2022, doi: 10.1049/ell2.12545.

[31] N. M. Ridler and M. Salter, "Measuring the capacitance coefficients of coaxial open-circuits with traceability to national standards," *IEEE Microwave Journal*, vol. 49, no. 10, pp. 138–154, 2006.